\documentclass[aps,prb,reprint,groupedaddress]{revtex4-1}
\usepackage{graphicx}
\usepackage{hyperref}
\usepackage{amsfonts}
\usepackage{amsmath,bm}
\usepackage{amssymb}
\usepackage{graphicx}
\usepackage{subfigure}
\usepackage{color}
\usepackage{float}

\draft
\begin{document}


\title{Hybrid photovoltaic and electron-tunneling converters}



\author{Shanhe Su$^{1}$}
\author{Jincan Chen$^{1}$}
\email[]{jcchen@xmu.edu.cn}
\author{Tien-Mo Shih$^{2}$}
\email[]{tshih111@gmail.com}

\affiliation{$1$.Fujian Key Laboratory of Semiconductor Materials and Applications and Department of Physics, Xiamen University, Xiamen 361005, People¡¯s Republic of China}
\affiliation{$2$.Institute for Complex Adaptive Matter, University of California, Davis, CA 95616, USA}
\date{\today}

\hspace{8mm}
\begin{abstract}
\textbf{Photon impingement is capable of liberating electrons in semiconductors. When the electron transport is primarily governed by temperature gradients, high irreversibilities will result, thus lowering converters' efficiencies. A fundamental study in the absence of photovoltaics\cite{1} has achieved the reduction of these irreversibilities by considering entropy changes due to electron flows. Here we present an unreported mechanism that integrates photovoltaic conversion and electron tunneling. Photon-excited electrons that occupy energy levels beyond windowed limits are first imprisoned inside the cathode, then given opportunities to rapidly re-thermalize, and eventually allowed to enter the tunnel. Energies wasted by both the irreversibility and the recombination are minimized with respect to the transmission energy and the transmission window that characterize the tunnel. Upon application of this mechanism to high-concentration solar cells, the proposed hybrid model outperforms others. It further provides a guide for elevating efficiencies in future photon-to-electron converters typified by third-generation photovoltaic systems.}
\end{abstract}

\maketitle


For photovoltaic cells, the Shockley-Queisser limit \cite{2,3,4}has long been recognized as the maximum theoretical efficiency. Several approaches have attempted to combine the photovoltaic and solar thermal technology to overcome this limit. For example, thermophotovoltaics \cite{5,6,7}is a direct process converting thermal energy to electricity via photon transports. It maneuvers to create photon emissions in a narrow wavelength range that is optimized for specific photovoltaic converters. Photon-enhanced thermionic emission solar cells utilize large densities of electron transmissions, thus elevating the conversion efficiency\cite{8,9}. Their theoretical limits are capable of reaching 45\% or higher at 1000-sun concentration\cite{8,10}. In the normal working condition, the model enjoys the merit of high temperatures, rendering it possible to jointly work with heat engines \cite{11}and thermoelectric generators \cite{12,13,14}.\par
In the thermionic mechanism, all electrons possessing energy levels higher than the potential barrier are allowed to exit the cathode, resulting in excessive electron-transport-related entropy production of the cathode-and-anode assembly. In addition, electrode plates are generally separated by a vacuum gap\cite{15,16}. Under high current densities, charge carriers may generate a space-charging regime within the vacuum\cite{17,18}, conceiving non-uniformity of electron distributions, which constitute high-energy barriers retarding electron movements. Here we propose a hybrid converter that minimizes wasted energies due to the photovoltaic recombination and the electron-exchange irreversibility. The quantum tunneling confines electron flows within a windowed energy range such that this minimization is achieved. If we equate probability distribution functions (PDFs) of the cathode and the anode at the electron-occupied energy level, $E^*$, the flow process is guaranteed to be reversible\cite{19,20}. When large temperature differences exist, the peak and full width at half maximum of PDFs will differ appreciably. Regardless of such differences, these two distributive curves will intersect, yielding a unique $E^*$ value. For example, if $T_C$=2000$K$, $T_A$=300$K$, $\mu_C$/$k_B$=1$K$, and $\mu_A$/$k_B$=2$K$, we obtain $E^*/k_B=2.1765K$, where $T_C$ and $T_A$ are cathode and anode temperatures; $\mu_C$ and $\mu_A$ cathode and anode chemical potentials; and $k_B$ the Boltzmann constant. This central idea lies in the optimization of the tunnel configuration to yield low irreversibilities. If the transmission window, $\Delta E$, narrows, few electrons are allowed to travel through the tunnel. For unnecessarily wide $\Delta E$, electron-occupied energy levels will deviate substantially from $E^*$. If the transmission energy, $E_o$, rises, the corresponding energy barrier will block electron flows. Conversely, at low $E_o$ values, a large number of overly hot electrons will reach the cold anode, thus increasing irreversibilities. Minimizing the radiative recombination and irreversibility, we manage to elevate the efficiency to 0.511 for 500-sun concentration.\par
The system schematic (Fig.1a) of a hybrid photovoltaic and electron-tunneling (HPET) converter consists of a solar concentrator, a photovoltaic cell, and a nanoscale vacuum-gap tunnel. The cathode and anode are fabricated with P-type boron-doped silicon and n-type silicon, respectively. They are separated by a vacuum-gapped tunnel, permitting electron tunneling upon the onset of a voltage bias. The system is operated in the steady state subject to negligible convective cooling.\par
Figure 1b describes the energy-band hierarchy diagram of HPET converter. Below-bandgap energy of solar photons is absorbed by a layer of the absorptive material coated on the cathode surface facing the sun, and is converted into the thermal energy. Under the assumption of unity transmittance for the coating layer with respect to above-bandgap energy, transmitted photons can excite electrons from the valence band to the conduction band of the cathode. Subsequently, these pumped electrons rapidly reach the thermal equilibrium, dictated by $T_C$ , and are distributed throughout the conduction band. Among them, those that carry energies within [$E_o$ - $\Delta E$/2, $E_o$ + $\Delta E$/2] are allowed to cross the tunnel, and reach the anode.\par
\begin{figure}[h]
\subfigure{\includegraphics[width=0.23\textwidth]{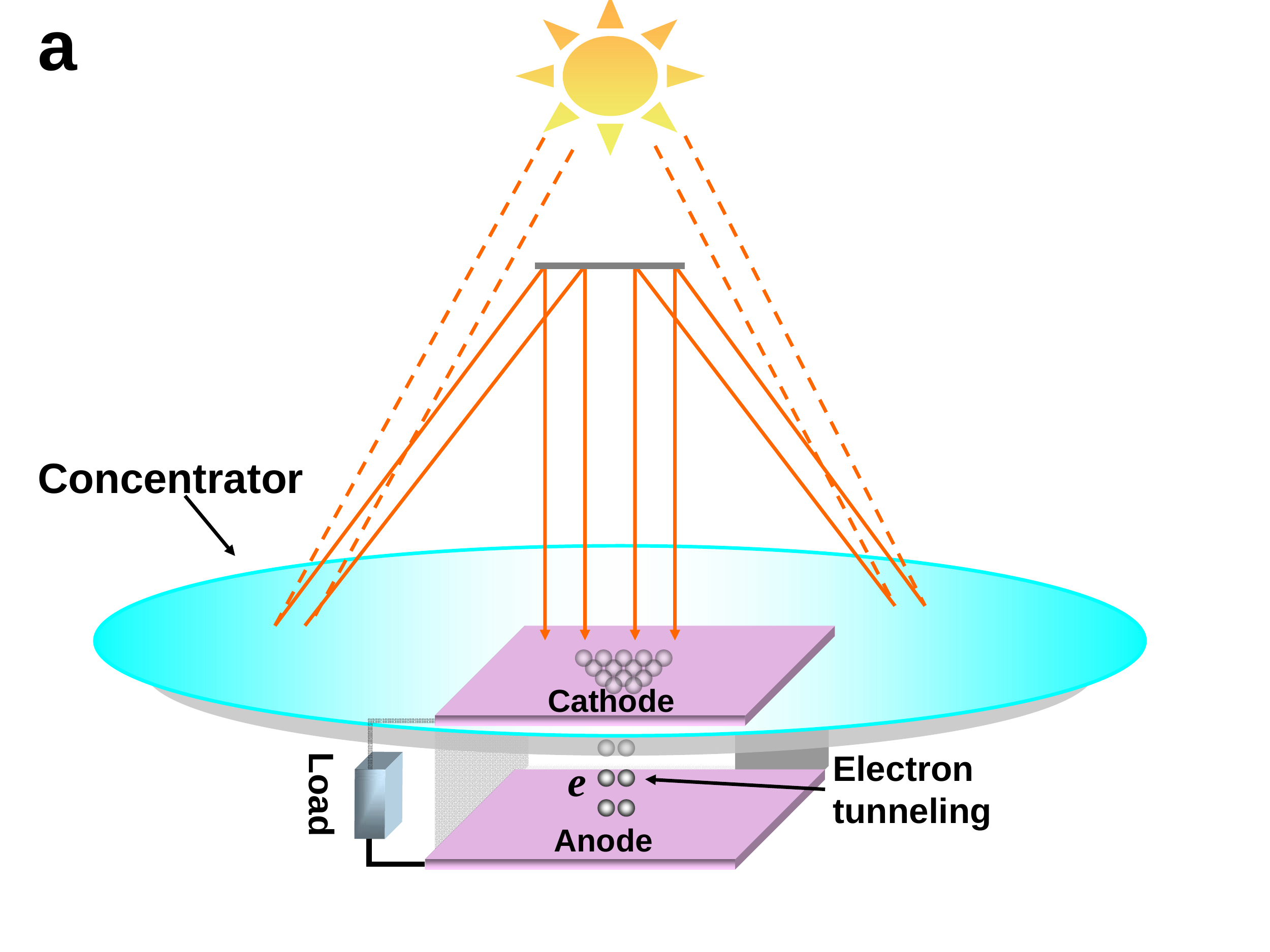}}
\subfigure{\includegraphics[width=0.23\textwidth]{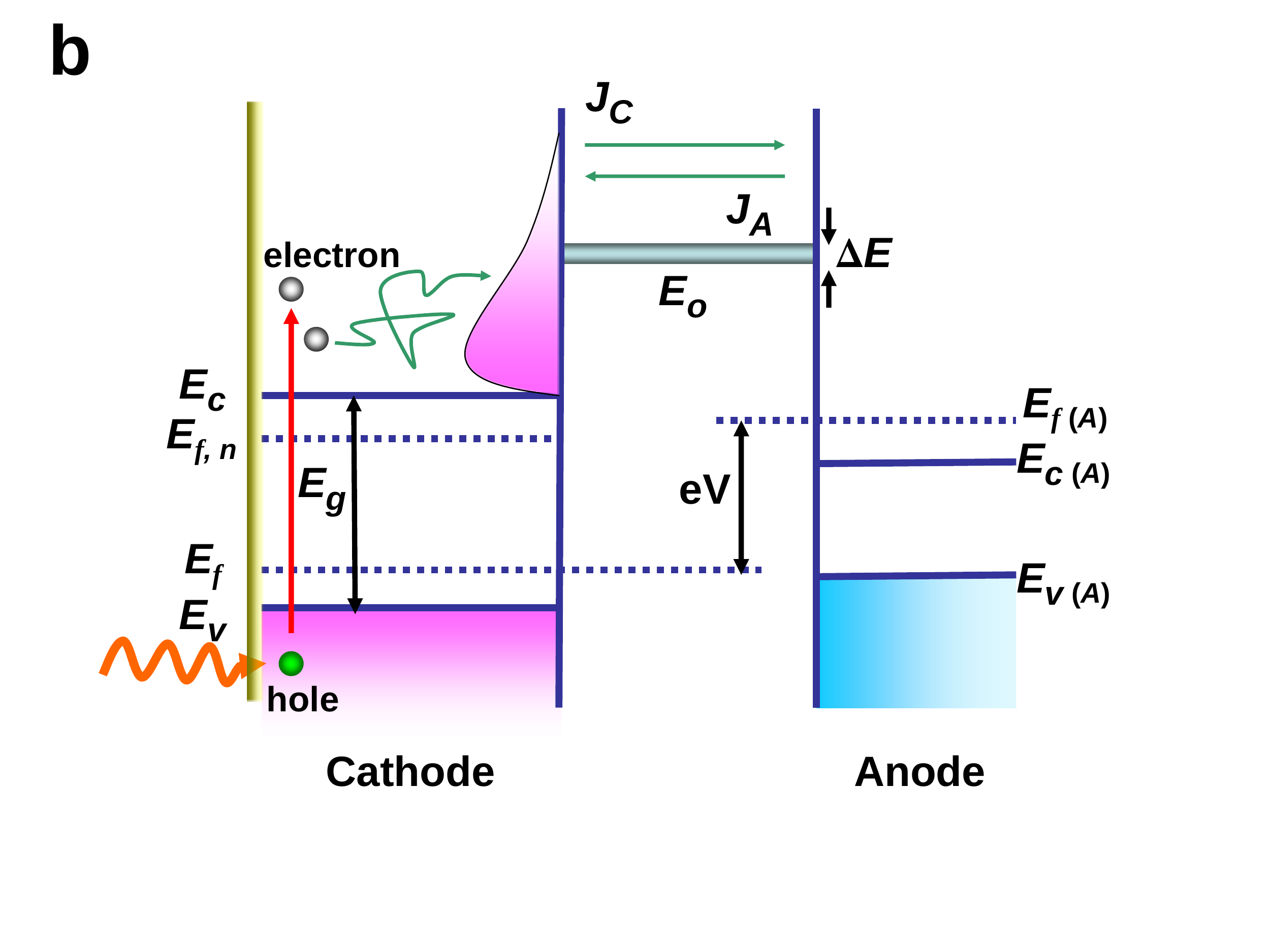}}
\caption{\textbf{The HPET process. $\bm a$,} System schematic of a hybrid photovoltaic and electron-tunneling (HPET) converter that consists of a solar concentrator, a photovoltaic cell, and a nanoscale vacuum-gap tunnel. Electrons are excited by impinging photons and travel through the tunnel from the cathode to the anode. {$\bm b$,} Band hierarchy diagram of HPET. In the cathode, $E_{\emph{f},n}$ denotes the quasi-Fermi level of photoexcited electrons; $E_\emph{f}$ the equilibrium Fermi level without photoexcitation; $E_\emph{g}$ the bandgap energy; $E_c$ the energy at the conduction-band minimum; and $E_v$ the energy at the valence-band maximum. Similarly, $E_{\emph{f}(A)}$, $E_{c(A)}$, and $E_{v(A)}$ are energy levels of the anode counterpart. Within the tunnel, $E_o$ denotes the transmission energy and $\Delta E$ is the transmission window. The operating voltage, $V$, is determined by $eV$ = $E_{\emph{f}(A)}$ - $E_\emph{f}$, where $e$ is the elementary positive charge. The net electrical current density equals $J_C$ - $J_A$ .}
\label{PIC1}
\end{figure}
In reference to Fig.1, the tunnel filters electron candidates. Only those whose energies, contributed by the streamwise-direction momenta, lie within the transmission window are allowed to enter the tunnel. The electrical current density flowing out of an electrode through the tunnel can be calculated by the Landauer equation\cite{21} as
\begin{equation}\label{1}
	\begin{split}
J=\frac{e}{4\pi^3}\int_{-\infty}^{\infty}\int_{-\infty}^{\infty}\int_{0}^{\infty}\emph{f}\,^{MB}
(E(\vec{k}),\mu,T)\\\upsilon(k_{x})\zeta(k_{x})dk_{x}dk_{y}dk_{z},
	\end{split}
\end{equation}
where $\emph{f}\,^{MB}(E(\vec{k}),\mu,T)$ = $\exp[-(E(\vec{k})-\mu)/(k_BT)]$ describes the distribution of electrons in the electrode at temperature $T$ and chemical potential $\mu$ according to the Maxwell-Boltzmann statistics;  $E(\vec{k})=\hbar^{2}(k^{2}_{x}+k^{2}_{y}+k^{2}_{z})/2m^{*}$ if the dispersion relation is parabolic; $k_{x}$, $k_{y}$ and $k_{z}$ wave vectors in a 3D space; $m^{*}$ the effective mass of electrons; $\upsilon(k_{x})=\hbar k_{x}/m^{*}$ the velocity in the streamwise direction; and $\zeta(k_{x})$ indicates the transmission probability of an electron traveling through the tunnel as a function of $k_{x}$. Equation (1) written in the energy form can be derived as $J=e/2\pi\int_{0}^{\infty}N(\mu,T)\zeta(E_{x})dE_{x}$, where $N(\mu,T)=[m^{*}k_{B}T/(\pi\hbar^3)]\exp[-(E_{x}-\mu)/(k_{B}T)]$ [Refs. 22, 23]. The net current density, $J_{net}$, of HPET converter is computed by the difference between the electrical current densities released by the cathode $J_{C}$ and the anode $J_{A}$. Namely,
\begin{equation}\label{2}
\begin{split}
  J_{net}=J_{C}-J_{A}=\frac{e}{2\pi}\int_{0}^{\infty}[N(E_{\emph{f},n},T_C)\\-N(E_{\emph{f}(A)},T_A)]\zeta(E_{x})dE_{x},
  \end{split}
\end{equation}
where, in non-degenerate semiconductors, $E_{\emph{f},n}=E_{\emph{f}}+k_{B}T_C\ln(n/n_{eq})$; $n_{eq}$ is the equilibrium electron concentration; and $n$ is the conduction band population of photoexcited electrons. Terms, $E_{\emph{f}}$ and $n_{eq}$, are computed using charge neutrality in the semiconductor (Supplementary S-1). The geometry of the tunnel is designed such that $E_o$ and $\Delta E$ are included in the $\zeta(k_{x})$ expression. This design is capable of minimizing energy losses caused by both the recombination in the cathode and the irreversibility during the electron transport.\par
We are now in the position to evaluate the performance of HPET converter under the concentrated sunlight by balancing electron generation, transport, and recombination in the cathode. Assumptions include: (1) cathode and anode plates are aligned in parallel, so that surface areas for the photon absorption, photon emission, and electron transmission are equal; (2) charge carriers' concentration and temperature are assumed to be uniform throughout the cathode; (3) the bottom of the cathode and the top of the anode are radiatively non-participating. Under the steady-state condition, the net rate of electrons flowing out of the cathode equals the difference between the rate of photon-electron-collision excitation ($G$) and that of photon-enhanced recombination, $R$, i.e.,
\begin{equation}\label{3}
  \emph{J}_{net}=eL(G-R),
\end{equation}
where $L$ is the thickness of the cathode. The generation rate of electrons, $G$, is computed using $G=\int_{0}^{\lambda_{g}}\Phi(\lambda)d\lambda$ , where  $\Phi(\lambda)$ is the spectral photon flux density and $\lambda_{g}$ is the wavelength corresponding to the bandgap energy $E_g$ of the semiconductor cathode. The concentrated AM1.5 direct circumsolar spectrum is used as the irradiance. The rate, $R$, can be expressed\cite{24} as $2\pi/(Lh^3c^2)\int_{E_g}^{\infty}(h\nu)^2/[\exp[h\nu/(k_{B}T_C)-1][np/(n_{eq}p_{eq})-1]d(h\nu)$, where $n$ and $p$ are concentrations of photoexcited electrons and holes (Supplementary S-2); $h\nu$ the photon energy; and $p_{eq}$ the equilibrium hole carrier concentration. Mechanisms of the Surface recombination\cite{25,26}, Auger recombination\cite{27,28}, and Shockley-Read-Hall recombination\cite{29,30} have been ignored herein.\par
The energy balance of the combined cathode and solar absorber is given by
\begin{equation}\label{4}
    P_{sun}=Q_{net}+P_{sa}+P_{rad}+P_{n,rad},
\end{equation}
where $Q_{net}$ is the net heat flux ($W/m^2$) exiting the cathode; $P_{sa}$ the radiation emitted from solar-absorbing material; $P_{rad}$ the equilibrium radiative recombination energy flux; $P_{n,rad}$ the photon-enhanced radiative re-combination energy flux (Supplementary S-3); and $P_{sun}$ the total power of the solar flux. Finally, the performance efficiency for the system is defined as
\begin{equation}\label{5}
    \eta=J_{net}V/P_{sun}.
\end{equation}\par
Figure 2 shows performances of the proposed system. Let us first examine the condition for small values of voltage $V$. Reverse current densities, $J_{A}$, from the anode to the cathode are suppressed because few electrons are capable of overcoming the energy barrier. Based on numerical simulations, we observe that $J_{C}$ remains independent of $V$, resulting in the constancy of $J_{net}$ (Fig. 2a). In the regime of large voltages, $J_{A}$ increases abruptly, thus leading to a drastic drop in $J_{net}$. In the vicinity of short circuit, $J_{net}$ increases as the transmission window, $\Delta E$, increases, because more electrons are capable of flowing through the tunnel as the transmission window widens. Conversely, open-circuit voltages decrease as $\Delta E$ increases. Next, optimal efficiencies are observed to prevail (Fig. 2b) for various $V$ and $\Delta E$. When $\Delta E$ is small, electrons are crowded in the cathode, resulting in an increase of the radiative recombination and a decrease of thermal energy carried by electrons leaving from the cathode to the anode.
In Fig. 2c, $\eta$ increases as $\Delta E$ increases in the small-$\Delta E$ regime ($\Delta E$ $<$ $\Delta E_m$), but it decreases as $\Delta E$ increases in the large-$\Delta E$ regime, where $\Delta E_m$ is the transmission window at the maximum efficiency. In the former, a surge of transporting electron results in a drop of radiative recombination. In the latter, the irreversibility significantly increases as $\Delta E$ increases. This increase rate exceeds the recombination rate, such that $\eta$ decreases after reaching its maximum.\par
Figure 2d depicts the optimization of $\eta$ at bandgap energy, $E_g = 1.1696 eV$. As seen in the figure, $\eta$ qualitatively peaks at $E_o = 2.1 eV$. In the small-$E_o$ regime, as $E_o$ increases, a great number of electrons are trapped in the cathode, resulting in both low $Q_{net}$ and high voltage, equivalently low irreversibilities. In the large-$E_o$  regime, as $E_o$ increases, by the same reason of a great number of electrons being trapped in the cathode, radiative recombination increases. It can be concluded that the sum of energies wasted by the irreversibility and the recombination is minimized with respect to the transmission energy and transmission window, i.e.,
\begin{equation}\label{6}
 \begin{cases}
   \partial(\psi_{ir}+\psi_R)/\partial E_o=0\\
    \partial(\psi_{ir}+\psi_R)/\partial\Delta E =0\\
  \end{cases},
\end{equation}
yielding optimal performance.\par
\begin{figure}
\subfigure{\includegraphics[width=0.23\textwidth]{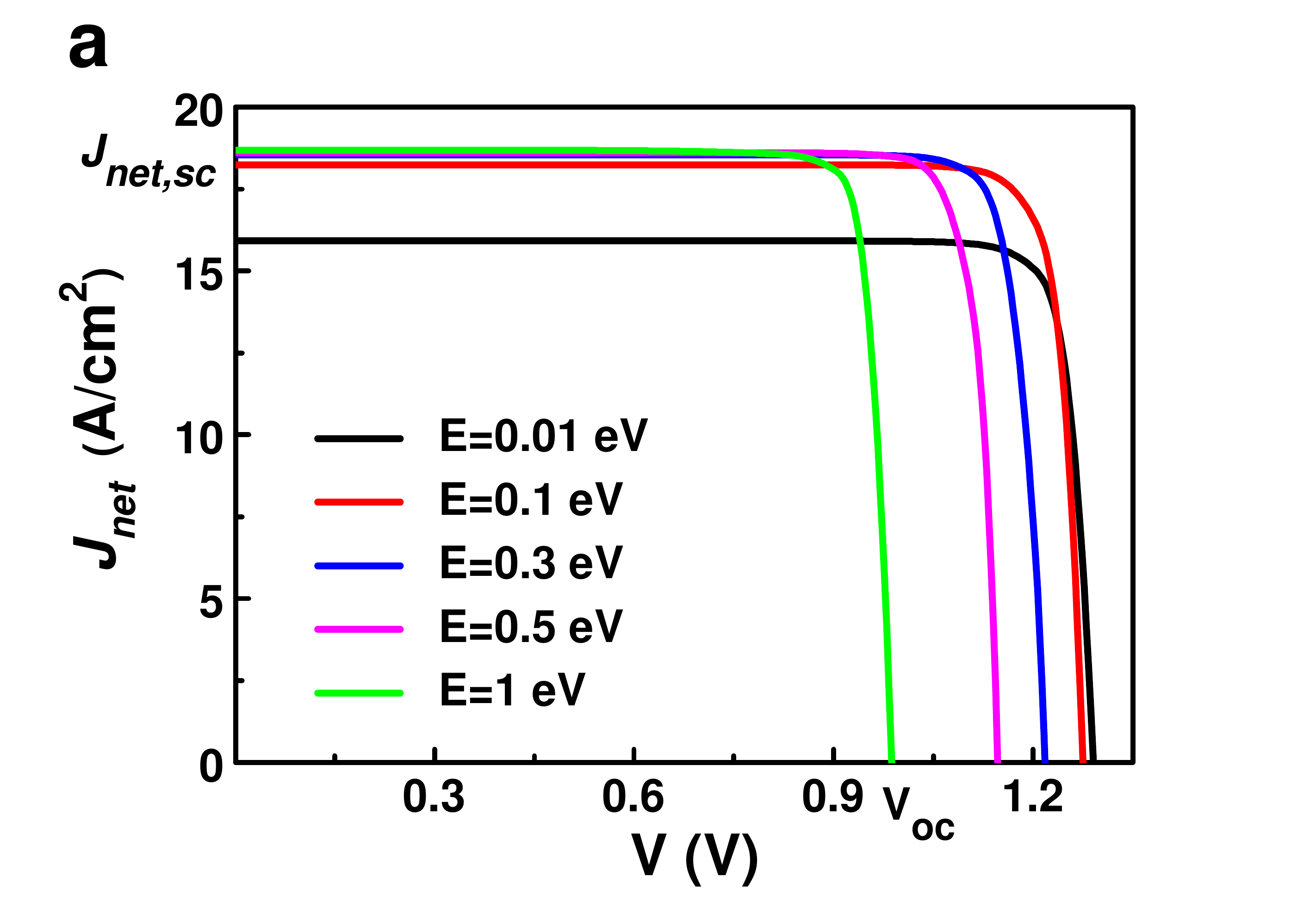}}
\subfigure{\includegraphics[width=0.23\textwidth]{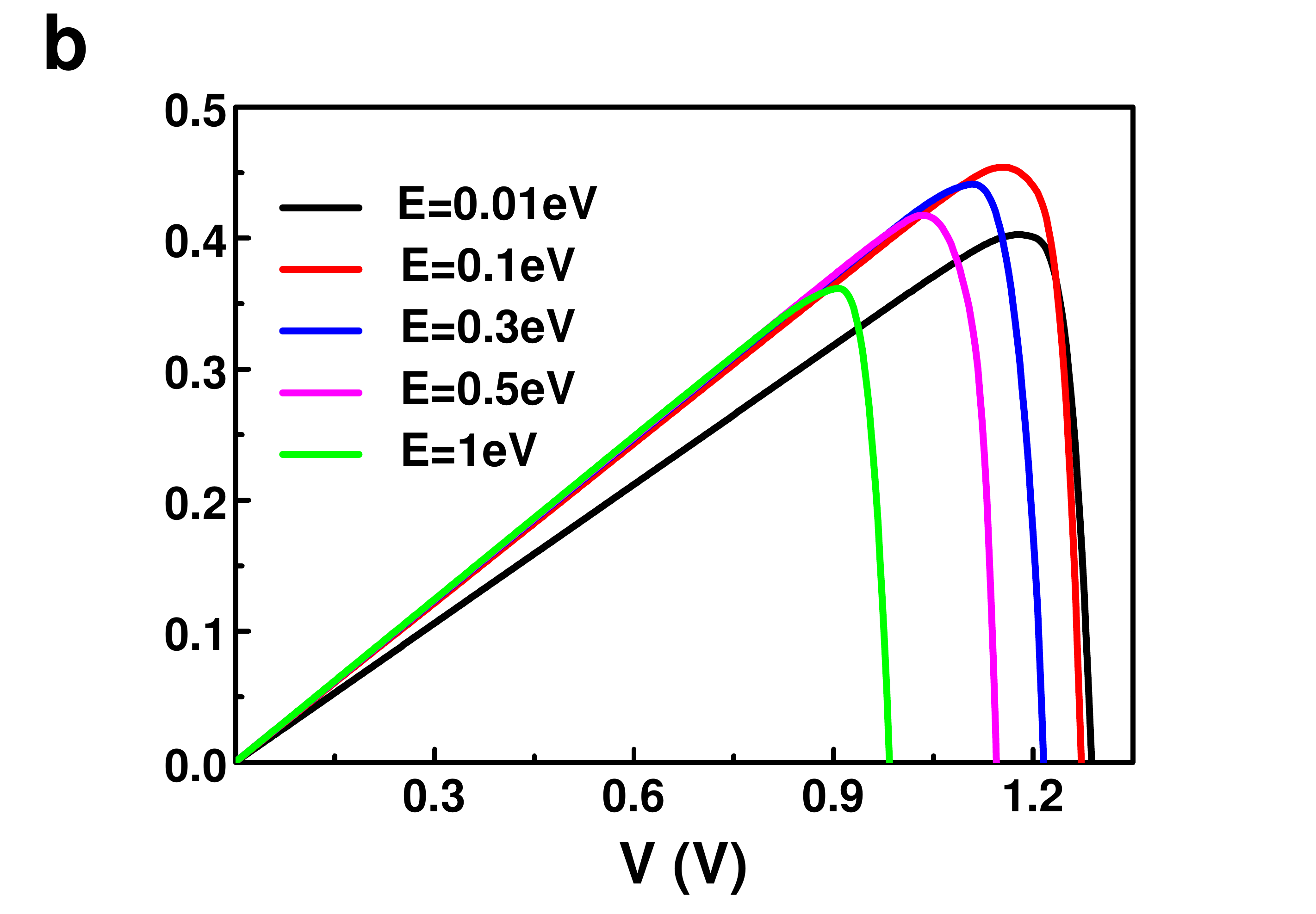}}
\subfigure{\includegraphics[width=0.29\textwidth]{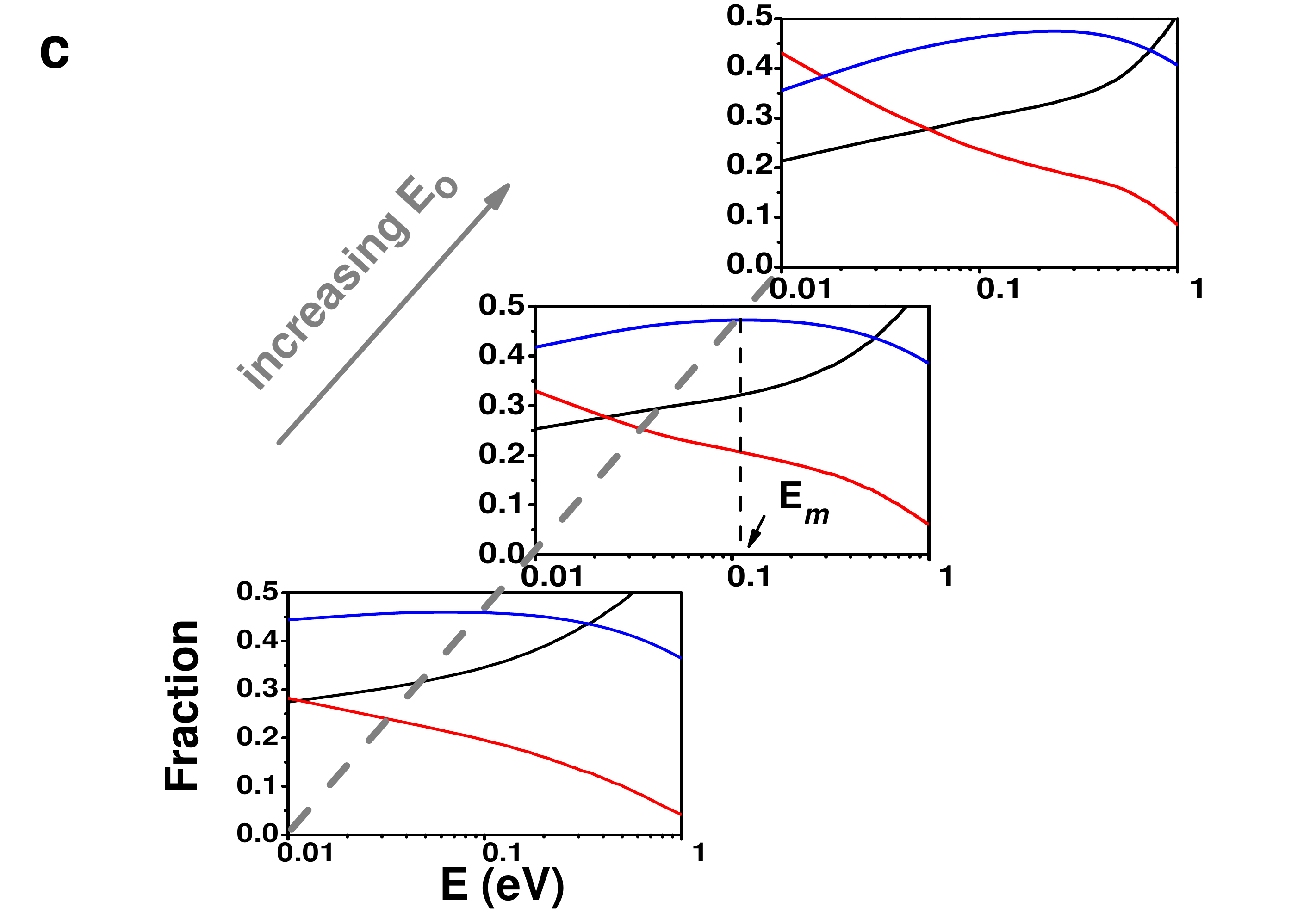}}
\subfigure{\includegraphics[width=0.29\textwidth]{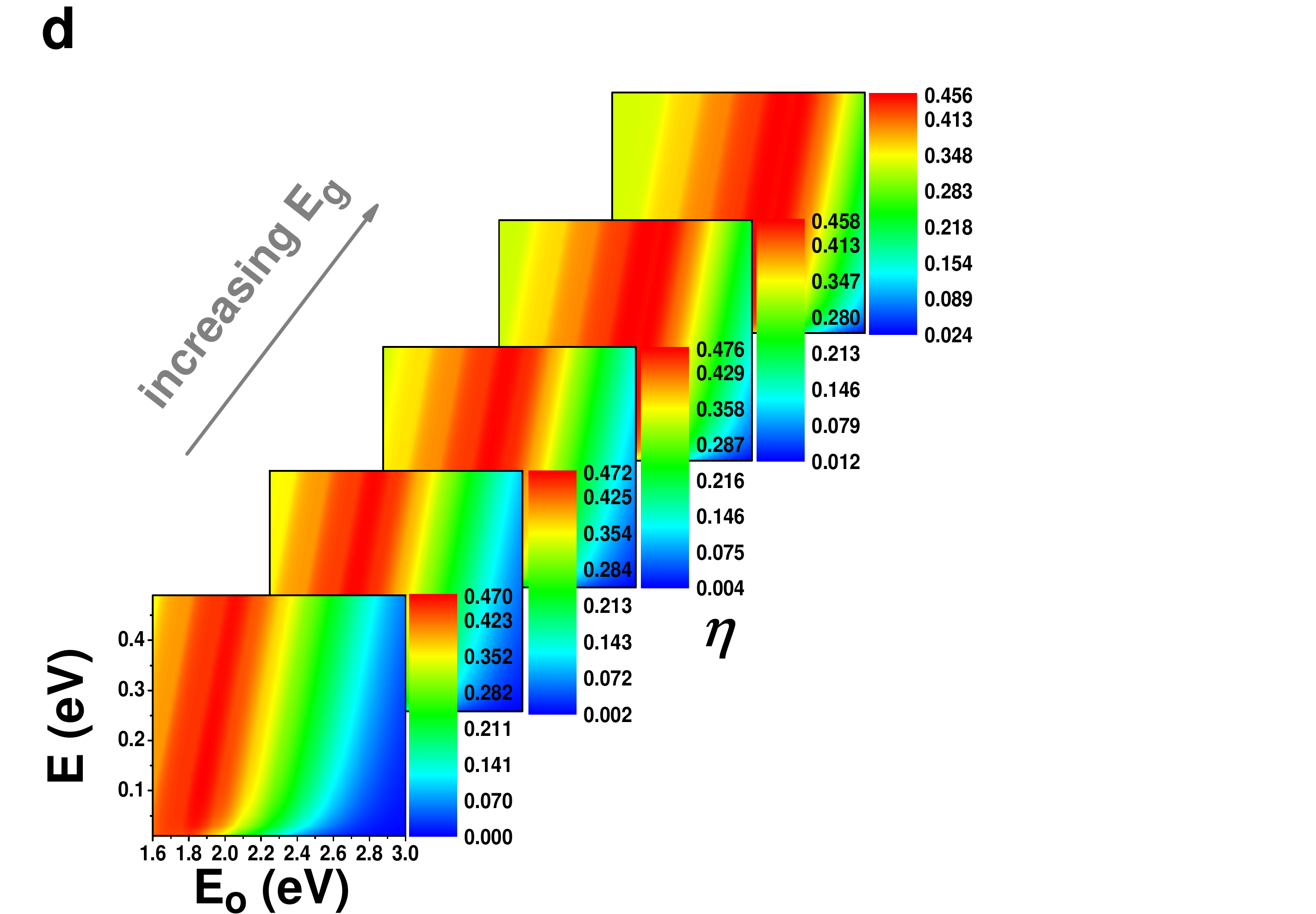}}
\caption{\textbf{The performance of HPET converter. $\bm a$,} Current density $J_{net}$ versus $V$. Subscripts $sc$ and $oc$ stand for ¡°short-circuit¡± and ¡°open-circuit¡±. $\bm b,$ Overall efficiency $\eta$ versus $V$. Both in $\bm a$ and $\bm b$, $E_o = 2.1eV$; $T_A = 500K$; $E_g = 1.1696eV$; and $C = 500$. These values are used unless otherwise mentioned specifically in the following discussion. $\bm c,$ Fraction of energy fluxes versus the transmission window, $\Delta E$, parametrized in $E_o $ = 2.0 eV, 2.1 eV, and 2.2 eV. Black lines represent the irreversibility loss, [$\psi_{ir}=(Q_{net}-J_{net}V)/P_{sun}$ ]; red lines represent the radiative recombination loss, [$\psi_R=(P_{sa}+P_{rad}+P_{n,rad})/P_{sun}$]; and blue lines represent the efficiency, $\eta$. $\bm d,$ Iso-efficiency contour plots versus two independent variables $\Delta E$ and $E_o$  with various $E_g$, whose values range from 0.9696 to 1.3696 with an increment equal to 0.1.}
\label{PIC2}
\end{figure}\par
\begin{figure}[H]
\centering
\subfigure{\includegraphics[width=0.32\textwidth]{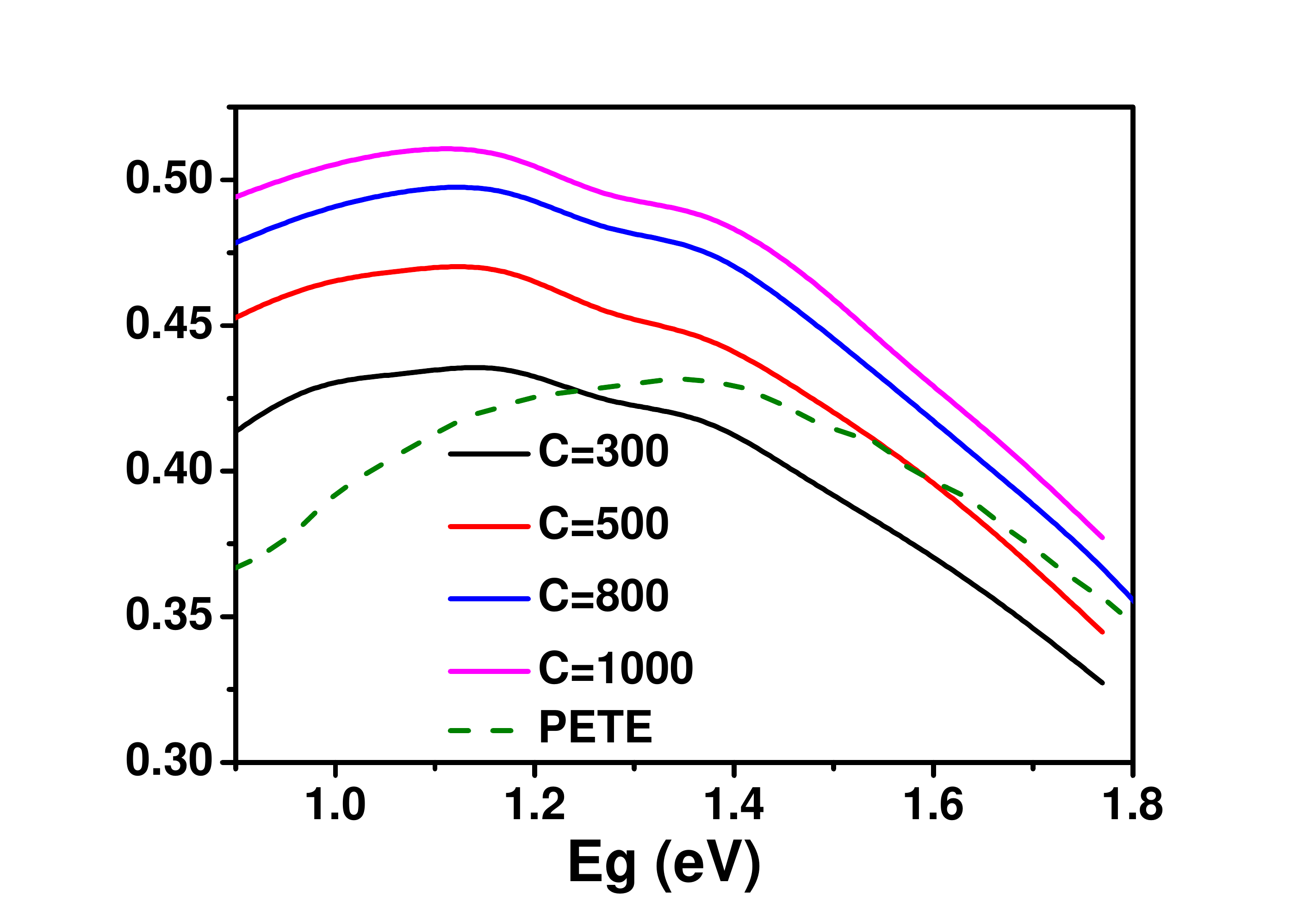}}
\caption{\textbf{Efficiency of HPET converter versus bandgap energy parametrized in the solar-flux concentration.} The transmission window $\Delta E$ = 0.1eV, and $E_o$ is optimized.}
\label{PIC3}
\end{figure}\par
Figure 3 shows $\eta$ versus $E_g$ parametrized in solar-flux concentration with optimal $E_g$ lying within [1.11 eV of Si, 1.4 eV of GaAs]. Benefits are pronounced, especially when the proposed model is jointly used with concentrators. By taking into account the closeness between PDFs in both electrodes, the irreversibility tends to decrease, yielding high efficiencies. Also shown is the dashed curve obtained by thermionic models\cite{8}. Efficiencies herein are depicted to be higher than those obtained by adopting the thermionic model with the solar-flux concentration, $C=1000$, for $E_g<1.5eV$. Only for $E_g>1.5eV$, $\eta$ of the proposed model with $C=500$ appears comparable with that of thermionic with $C=1000$. In general, as the concentration increases, $\eta$ increases monotonically.
Finally, in Fig.4, to examine the applicability of the proposed model to various shapes of the transmission window, we replace the ideal rectangle with a Gaussian distribution described as $\zeta(k_{x})=\exp[-(E_x-E_o)^2/w]$, where $E_o$ is the energy level of the peak and $\emph{w}$ is the width-like parameter (Fig. 4a). In comparison with the ideal rectangular window, now electrons carry energies that deviate farther from $E_o$, resulting in higher irreversibilities. Overall, efficiencies (Fig. 4b) are similar to, but slightly lower than, those shown in Fig. 3.\par
\begin{figure}
\subfigure{\includegraphics[width=0.25\textwidth]{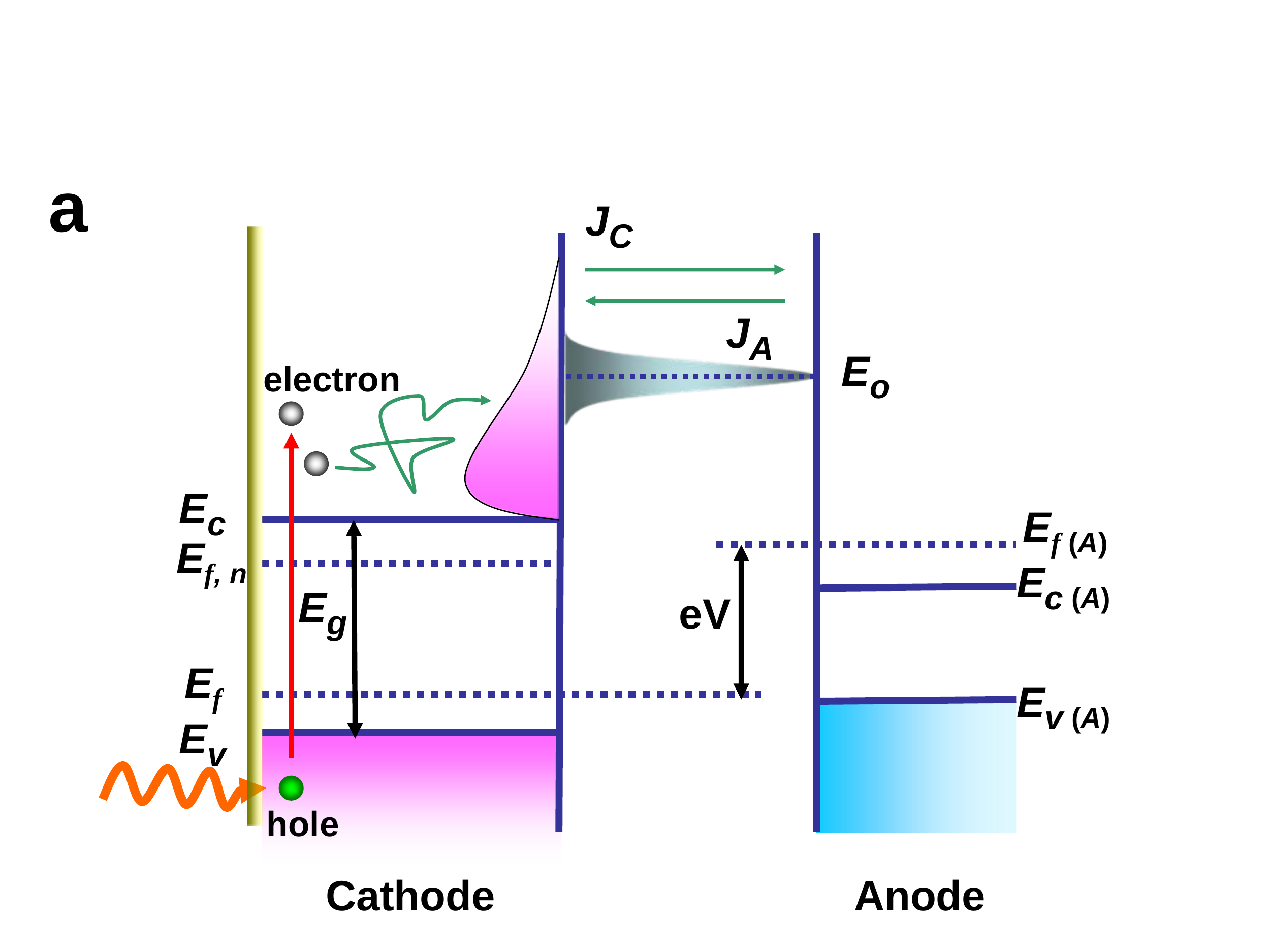}}
\subfigure{\includegraphics[width=0.22\textwidth]{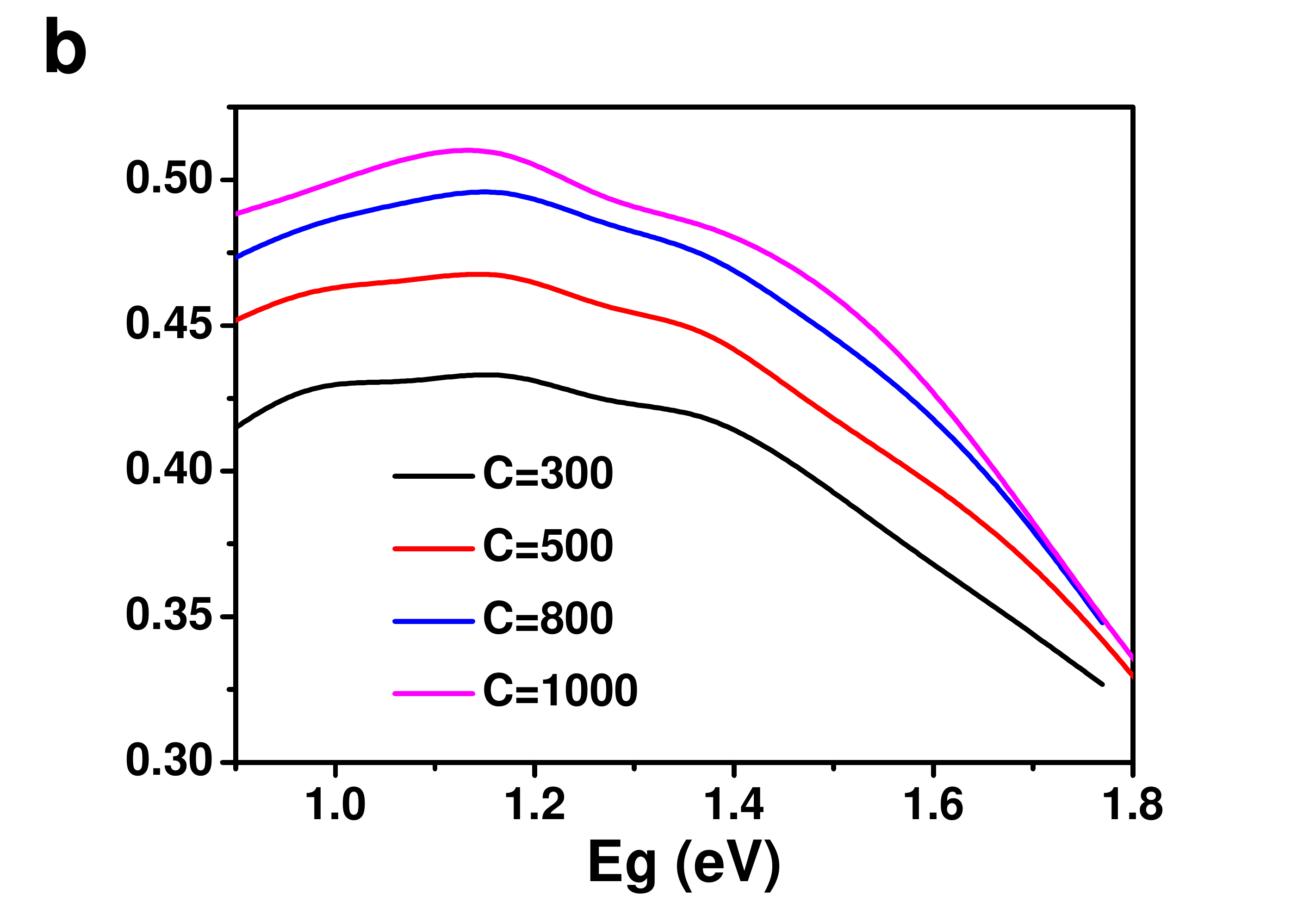}}
\caption{\textbf{HPET converter with a semi-ideal transmission window. $\bm a$,} Band hierarchy diagram employing a Gaussian transmission window. {$\bm b$,} Efficiency varying with $E_g$ parametrized in the solar-flux concentration. The parameter $\emph{w}$ equals 0.01eV, and $E_o$ has been optimized.}
\label{PIC4}
\end{figure}\par


\begin{thebibliography}{00}


\bibitem{1}
~Humphrey, T.E., Newbury, R., Taylor, R.P. $\&$ Linke, H. {Reversible quantum Brownian heat engines for electrons. } {\em Phys. Rev. Lett.} \textbf{89},
  ~116801 (2002).

\bibitem{2}
~Shockley, W. $\&$ Queisser, H.J. {Detailed balance limit of efficiency of p-n junction solar cells.} {\em J. Appl. Phys.} \textbf{32},
  ~510-519 (1961).

\bibitem{3}
~Krogstrup, P. et al. {Single-nanowire solar cells beyond the Shockley-Queisser limit.} {\em Nature Photon.} \textbf{7},
  ~306-310 (2013).

\bibitem{4}
~Polman, A. $\&$ Atwater, H.A. {Photonic design principles for ultrahigh-efficiency photovoltaics.} {\em Nature mater.} \textbf{11},
  ~174-177 (2012).

\bibitem{5}
~Lenert, A. et al. {A nanophotonic solar thermophotovoltaic device.} {\em Nature Nanotech.} \textbf{9},
  ~126-130 (2014).

\bibitem{6}
~Chan, W. R. et al. {Toward high-energy-density, high-efficiency, and moderate-temperature chip-scale thermophotovoltaics.} {\em Proc. Natl. Acad. Sci.} \textbf{110},
  ~5309-5314 (2013).

\bibitem{7}
~Fraas, L.M., Avery, J.E. $\&$ Huang, H.X. {Thermophotovoltaic furnace generator for the home using low bandgap GaSb cells.} {\em Semicond. Sci. Technol.} \textbf{18},
  ~S247-S253 (1997).

\bibitem{8}
~Schwede, J.W. et al. {Photon-enhanced thermionic emission for solar concentrator systems.} {\em Nature Mater.} \textbf{9},
  ~762-767 (2010).

\bibitem{9}
~Schwede, J.W. et al. {Photon-enhanced thermionic emission from heterostructures with low interface recombination.} {\em Nature commun.} \textbf{4},
  ~1576 (2013).

\bibitem{10}
~Segev, G., Rosenwaks, Y. $\&$ Kribus, A. {Effciency of photon enhanced thermionic emission solar converters.} {\em Sol. Energy Mater. Sol. Cells} \textbf{107},
  ~125-130 (2012).

\bibitem{11}
~Segev, G., Kribus, A. $\&$ Rosenwaks, Y. {High performance isothermal photo-thermionic solar converters.} {\em Sol. Energy Mater. Sol. Cells} \textbf{113},
  ~114-123  (2013).

\bibitem{12}
~Su, S., Wang, Y., Wang, J., Xu, Z. $\&$ Chen, J. {Material optimum choices and parametric design strategies of a photon-enhanced solar cell hybrid system.} {\em Sol. Energy Mater. Sol. Cells} \textbf{128},
  ~112-118 (2014).

\bibitem{13}
~Bell, L.E. {Cooling, heating, generating power, and recovering waste heat with thermoelectric systems.} {\em Science} \textbf{321},
  ~1457-1461 (2008).

\bibitem{14}
~Karni, J. {Solar energy: The thermoelectric alternative.} {\em Nature Mater.} \textbf{10},
  ~481-482 (2011).

\bibitem{15}
~Hatsopoulos, G. N. $\&$ Gyftopoulos, E. P. {\em Thermionic energy conversion} (Cambridge, 1979).

\bibitem{16}
~Lee, J. H., Bargatin, I., Melosh, N. A. $\&$ Howe, R. T. {Optimal emitter-collector gap for thermionic energy converters.} {\em Appl. Phys. Lett.} \textbf{100},
  ~173904 (2012).

\bibitem{17}
~Ito, T. $\&$ Cappelli, M.A. {Optically pumped cesium plasma neutralization of space charge in photon-enhanced thermionic energy converters.} {\em Appl. Phys. Lett.} \textbf{101},
  ~213901 (2012).

\bibitem{18}
~Su, S., Wang, Y., Liu, T., Su, G. $\&$ Chen, J. {Space charge effects on the maximum efficiency and parametric design of a photon-enhanced thermionic solar cell.} {\em Sol. Energy Mater. Sol. Cells} \textbf{121},
  ~137-143 (2014).

\bibitem{19}
~Humphrey, T.E. $\&$ Linke, H. {Reversible thermoelectric nanomaterials.} {\em Phys. Rev. Lett.} \textbf{94},
  ~096601 (2005).

\bibitem{20}
~O'Dwyer, M.F., Humphrey, T.E. $\&$ Linke H. {Concept study for a high-efficiency nanowire based thermoelectric.} {\em Nanotechnology} \textbf{17},
  ~S338-S343 (2006).

\bibitem{21}
~Davies, J.H. {\em The physics of low-dimensional semiconductors: an introduction} (Cambridge university press, 1998).

\bibitem{22}
~O'Dwyer, M.F., Lewis, R.A., Zhang, C. $\&$ Humphrey, T.E. {Electronic efficiency in nanostructured thermionic and thermoelectric devices.} {\em Phys. Rev. B} \textbf{72},
  ~205330 (2005).

\bibitem{23}
~O'Dwyer, M.F., Humphrey, T.E., Lewis, R.A., $\&$ Zhang, C. {Efficiency in nanometre gap vacuum thermionic refrigerators.} {\em J. Phys. D: Appl. Phys.} \textbf{42},
  ~035417 (2009).

\bibitem{24}
~W\"{u}rfel, P. {\em Physics of solar cells: From basic principles to advanced concepts 2nd edn} (Wiley-VCH, 2009).

\bibitem{25}
~Oh, J., Yuan, H.C. $\&$ Branz, H.M. {An 18.2\%-efficient black-silicon solar cell achieved through control of carrier recombination in nanostructures.} {\em Nature Nanotech.} \textbf{7},
  ~743-748 (2012).

\bibitem{26}
~Atwater, H. A. $\&$ Polman, A. {Plasmonics for improved photovoltaic devices.} {\em Nature Mater.} \textbf{9},
  ~205-213 (2010).

\bibitem{27}
Javaux, C. et al. {Thermal activation of non-radiative Auger recombination in charged colloidal nanocrystals.} {\em Nature Nanotech.} \textbf{8},
  ~206-212 (2013).

\bibitem{28}
Yu, L. $\&$ Zunger, A.  {Identification of potential photovoltaic absorbers based on first-principles spectroscopic screening of materials.} {\em Phys. Rev. Lett.} \textbf{108},
  ~068701 (2012).

\bibitem{29}
~Assmann, E. et al. {Oxide heterostructures for efficient solar cells.} {\em Phys. Rev. Lett.} \textbf{110},
  ~078701 (2013).

\bibitem{30}
~Cowan, S. R., Roy, A. $\&$ Heeger, A. J. {Recombination in polymer-fullerene bulk heterojunction solar cells.} {\em Phys. Rev. B.} \textbf{82},
  ~245207 (2010).

\end{thebibliography}
\end{document}